\documentclass[aps,pre,twocolumn,nofootinbib,superscriptaddress,showpacs,showkeys]{revtex4-1}
\usepackage{color}
\usepackage{xcolor}

\usepackage{amssymb}
\usepackage{amsmath}
\usepackage{bm}
\usepackage{graphicx}
\usepackage{epstopdf}
\usepackage{newtxtext,newtxmath}
\usepackage{xfrac}
\usepackage{amsfonts}
\usepackage{CJK}
\usepackage{soul}
\usepackage[normalem]{ulem}

\definecolor{dkgreen}{rgb}{0,0.6,0}
\definecolor{gray}{rgb}{0.5,0.5,0.5}
\definecolor{mauve}{rgb}{0.58,0,0.82}

\def\BigRoman{\uppercase\expandafter{\romannumeral\number\count 255 }}
\def\Romannumeral{\afterassignment\BigRoman\count255=}

\begin{document}


\setcounter{page}{1}
\title{Contagion dynamics on hypergraphs with nested hyperedges}
\author{Jihye Kim}
\affiliation{Department of Physics, Korea University, Seoul 02841, Korea}
\author{Deok-Sun Lee}
\email{deoksunlee@kias.re.kr}
\affiliation{School of Computational Sciences and Center for AI and Natural Sciences, Korea Institute for Advanced Study, Seoul 02455, Korea}
\author{K.-I.~Goh}
\email{kgoh@korea.ac.kr}
\affiliation{Department of Physics, Korea University, Seoul 02841, Korea}
\date{\today}
\begin{abstract}
In complex social systems encoded as hypergraphs, higher-order ({\it i.e.}, group) interactions taking place among more than two individuals are represented by hyperedges.
One of the higher-order correlation structures native to hypergraphs is the nestedness: Some hyperedges can be entirely contained (that is, nested) within another larger hyperedge, which itself can also be nested further in a hierarchical manner. Yet the effect of such hierarchical structure of hyperedges on the dynamics has remained unexplored. 
In this context, here we propose a random nested-hypergraph model with tunable level of nestedness and investigate the effects of nestedness on a  higher-order susceptible-infected-susceptible process. 
By developing an analytic framework called the facet approximation, 
we obtain the steady-state fraction of infected nodes on the random nested-hypergraph model more accurately than existing methods.
Our results show that the hyperedge-nestedness affects the phase diagram significantly. Monte Carlo simulations support the analytical results. 
\end{abstract}
\maketitle
\section{\label{sec:level1}Introduction}
Understanding emergent phenomena \cite{r1} on networks is a fundamental subject of research in network science. In particular, the description of spreading processes has been one of the most central fields in complex networks, such as the propagation of epidemic diseases \cite{r2} and rumors, adoption of innovation, opinion formation, and many more \cite{newman1,goh}. Network models are  indispensable for quantitative analysis of the contagion dynamics. However, a large body of research has neglected higher-order interactions among more than two nodes. Because myriad systems \cite{bioche,sim_col2,eco1,eco2,trade} are prolific in higher-order interactions, many researchers have turned their attention to hypergraphs and simplicial complexes for expanding the paradigm of pairwise interactions into that of higher-order interactions \cite{physicsreport,blue,bianconi2021,pink}. 

Two major structural models for higher-order interactions are hypergraphs and simplicial complexes. 
A hypergraph consists of nodes and hyperedges. Nodes represent the individual constituents; hyperedges are the higher-order, group-wise interactions among the nodes. Any set of (two or more) nodes can be `connected' by a hyperedge. In contrast, in a simplicial complex the group-wise interaction is represented by the simplex. A simplicial complex  is defined as a set of simplices, with an extra condition that every face of a simplex also exists within it. In this paper, we will use the hypergraph as the main theoretical tool. 

\begin{figure}[b]
\centering
\includegraphics[width=.95\linewidth]{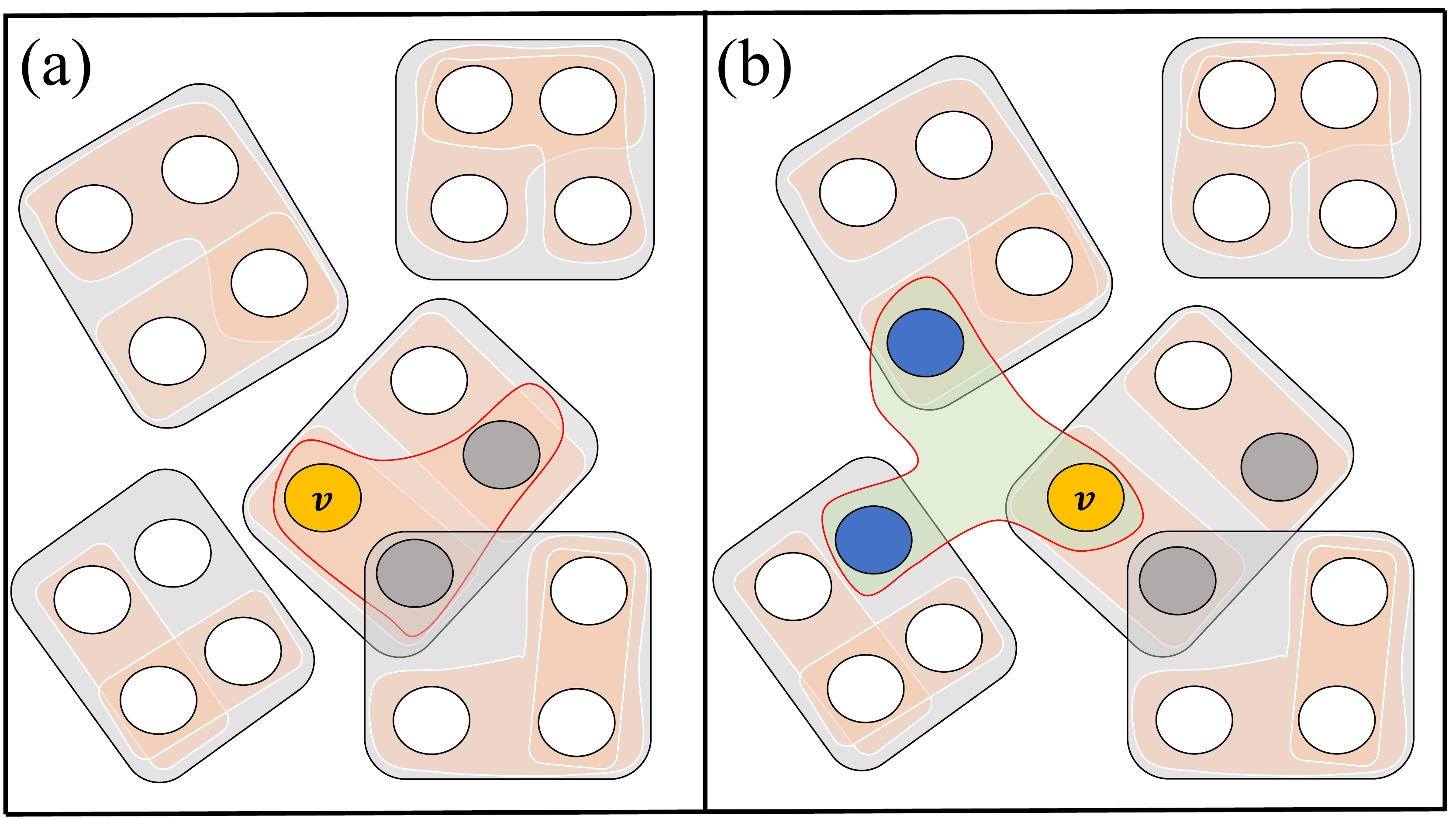}
\caption{(a) An example of hypergraph with nested hyperedges. In this example, red-colored hyperedges are nested within gray-colored hyperedges. 
(b) An example of hypergraph obtained from (a) by random-rewiring the red-contoured hyperedge. 
}
    \label{fig:rewiring}
\end{figure}

In higher-order interacting complex systems, a greater spectrum of native higher-order correlation structures of group interactions emerge, such as the higher-order component \cite{jungho} and the hypergraph core \cite{kq}. In this paper we focus on the {\it nestedness}: Some hyperedges can be entirely contained (that is, nested) within another larger hyperedge, which itself can also be nested further in a hierarchical manner (see, for example, Fig.~\ref{fig:rewiring}(a), where red-colored hyperedges are nested within gray-colored ones). Nestedness is observed ubiquitously~\cite{motif}. In a social network, we often witness small groups of close friends within a larger community, for instance. Such a hierarchical nested structure creates strong correlations in the dynamical processes over it and thus is expected to impact the collective dynamics.  
Yet, the effects of hierarchical nested structure of hyperedges on higher-order dynamic processes remained largely overlooked, despite a recent surge of works  \cite{SCM,r26,r28,r29,hypergraph,r27,st2021social,universal,cca,st2022influential,jungho} to study the interplay between hypergraph structure and dynamics. 

The main objective of this paper is to study the effect of hyperedge-nestedness on higher-order contagion dynamics. To this end, we 
first propose a model of random nested-hypergraphs with tunable level of nestedness in Sec.~II. 
We employ a higher-order susceptible-infected-susceptible (SIS) process \cite{st2022influential} as an archetypal contagion process,
and formulate an analytical approximation framework called the facet approximation in Sec.~III.  
We compute the steady-state fraction of infected nodes on the random nested-hypergraph model for different level of nestedness and investigate how the hyperedge-nestedness affects the critical behaviors in Sec.~IV. 
Finally, we discuss the main results and present the conclusions in Sec.~{\Romannumeral 5}.

\section{Random nested-hypergraph model}
To address the impact of hyperedge-nestedness systematically, it is desirable to work with the hypergraph with tunable level of nestedness. To motivate, let us note that structure-wise a simplicial complex is fully-nested because each maximal simplex, called the facet, contains every possible smaller simplices within it; a (sparse) random hypergraph, on the other hand, is hardly nested. Based on these observations, we define the random nested-hypergraph model with tunable nestedness in the following way. The hypergraph consists of $N$ nodes.

\begin{itemize}
\item[$(i)$] Make $H_{s_m}$ hyperedges of size $s_m$: For each, select $s_m$ distinct nodes randomly and connect them by a hyperedge. 
Throughout the model, we do not allow multi-hyperedges: If there exists an hyperedge with the identical node set already, it is rejected and re-tried. The hyperedges made in this step play the role of facets in a simplicial complex.

\item[$(ii)$] For each hyperedge created in step (i), every proper subset of cardinality greater than one also becomes a hyperedge. All the hyperedges made in this step are {nested} at this stage.

\item[$(iii)$] For each hyperedge of size $s$ created in step (ii), rewire it into a new hyperedge with probability $1-\varepsilon_s$. 
The rewiring is done by first selecting a random node $v$ as a pivot node and then replacing the remaining $(s-1)$ nodes in that hyperedge by randomly selected $(s-1)$ nodes outside the pivot node's facet (see Fig.~\ref{fig:rewiring}). Note that multi-hyperedges are forbidden.
\end{itemize}
The parameter $\varepsilon_s$ is referred to as the hyperedge-nestedness parameters and controls the global level of nestedness. After steps (i) and (ii), a fully-nested hypergraph is obtained, which is structurally equivalent to a simplicial complex except for the exclusion of 0-simplex. As one rewires more, the level of nestedness decreases. If one rewires all the nested hyperedges, the resulting hypergraph becomes essentially random. 
In that sense, the random nested-hypergraph model can be thought to interpolate between simplicial complexes and random hypergraphs. 

Before proceeding, let us compute some quantities to be used for the analytic calculations later. 
We assume throughout that $H_{s_m}={\cal O}(N)$ and $s_m={\cal O}(1)$, so that the chance of two distinct hyperedges of size $s_m$ sharing more than one node is negligible.
First, the expected number of the hyperedges of size $s$ remaining nested after the rewiring process, denoted by $\langle H^{(\mathrm{nested})}_s \rangle$ is given by $\langle H^{(\mathrm{nested})}_s \rangle=\varepsilon_s H_{s_m}{s_m \choose s}$. Here ${a \choose b}\equiv \frac{a!}{(a-b)!b!}$. 
The number of nested (rewired) hyperedges of size $s$ accommodating a node $i$ is denoted as $k^{(s,\mathrm{nested})}_i$ ($k^{(s,\mathrm{rewired})}_i$). Their expected values $\langle k^{(s,\mathrm{nested})}_i \rangle$ and $\langle k^{(s,\mathrm{rewired})}_i \rangle$ depend on $k^{(s_m)}_i$ denoting the number of hyperedges of size $s_m$ to which the node $i$ belongs and are given as follows:
\begin{align}
\langle k^{(s,\mathrm{nested})}_i \rangle&=k^{(s_m)}_i{s_m-1 \choose s-1}\varepsilon_s~,\nonumber
\\
\langle k^{(s,\mathrm{rewired})}_i \rangle &=k^{(s_m)}_i{s_m-1 \choose s-1}(1-\varepsilon_s)\frac{1}{s}\nonumber
\\
&\quad+(H_{s_m}-k^{(s_m)}_i){s_m \choose s}(1-\varepsilon_s)\frac{s-1}{N-s_m}\nonumber
\\
&\approx (1-\varepsilon_s){s_m-1 \choose s-1}\left[\frac{k^{(s_m)}_i}{s}+\frac{m_{s_m}(s-1)}{s}\right],
\label{eq:averdeg}
\end{align}
with $m_{s}\equiv s H_{s_m}{s_m \choose s}/N$. Here we further assume that $H_{s_m}\gg k^{(s_m)}_i$. In Eq.~(1), $k^{(s_m)}_i{s_m-1 \choose s-1}$ is the number of nested hyperedges of size $s$ to which a node $i$ belongs before step $(iii)$; the fraction $\varepsilon_s$ of them remain nested after step $(iii)$, on average. The other fraction $1-\varepsilon_s$ may take $i$ as the pivot with probability $1/s$. $(H_{s_m}-k^{(s_m)}_i){s_m \choose s}(1-\varepsilon_s)$ is the number of to-be rewired hyperedges of size $s$ originally from other hyperedges of size $s_m$ to which $i$ does not belong; the node $i$ can be selected with probability $(s-1)/(N-s_m)$.
We refer to the vector $\vec{\mathbf{k}}_i=\left[ k^{(2,\mathrm{nested})}_i, k^{(2,\mathrm{rewired})}_i,...,k^{(s_m-1,\mathrm{nested})}_i,k^{(s_m-1,\mathrm{rewired})}_i,k^{(s_m)}_i\right]$ as the \textit{degree vector} of node $i$ \cite{r28}.
We denote the degree-vector distribution by $P_{\vec{\mathbf{k}}}=\sum\limits_{i} \delta_{\vec{\mathbf{k}}_i,\vec{\mathbf{k}}}/N$ with  $\vec{\mathbf{k}}\equiv\left[k^{(2,\mathrm{nested})},k^{(2,\mathrm{rewired})},...,k^{(s_m-1,\mathrm{nested})},k^{(s_m-1,\mathrm{rewired})},k^{(s_m)}\right]$. 

\section{Facet approximation for higher-order SIS processes}
\subsection{Higher-order SIS processes}
In this study, we consider the higher-order SIS process \cite{st2022influential}, as archetypal contagion dynamics.  Nodes can be in either S (susceptible) or I (infected) state. The dynamics is composed of the following two processes:

\begin{itemize}
\item[$(i)$] For each hyperedge of size $s$ with $n$ ($<s$) infected nodes, 
a susceptible node in the hyperedge catches a disease at rate $\beta(s,n)$, which is a function of $s$ and $n$.

\item[$(ii)$] Every infected node in the hypergraph independently turns back into the susceptible state  at rate $\mu$.
\end{itemize}

\subsection{Facet approximation} 
In networks, various analytical approaches to approximate the relationship between the order parameter (such as the density of infectious nodes, the consensus value, {\it etc.}) and stochastic node-level interactions in the dynamic processes were developed \cite{binary}. Existing approximation frameworks, which do not depend on specific realized networks, may be classified into three schemes \cite{r35}: the mean-field approximation (MFA), the pair approximation (PA), and the clique approximation (CA). 
Notably, the CA framework of Ref.~\cite{propagation} attempts to capture dynamical correlations in networks at a clique-level.

The study of higher-order dynamic processes on hypergraphs thus far has mostly used the generalizations of MFA and PA ~\cite{SCM,r26,r28}. In the presence of higher-order correlation structures like nestedness, however, its utility is limited ~\cite{r26}. There are attempts to take the `simplicial' structure into account ~\cite{r29,kuramoto}, yet applicable only to the `simplicial' {\it viz.} fully-nested case. 

To fill this gap, here we introduce and formulate an analytical framework that can be used for hypergraphs with arbitrary nestedness, which we call the facet approximation (FA). 
The main idea of FA is to distinguish the infections through the nested hyperedges and those through free, non-nested ones. 
For the former strong correlations due to the nestedness must be taken into account for accuracy. In FA, this correlation is accounted for by approximating the infections through nested hyperedges by means of the local mean-field associated with the facet, hence its name FA. It can be regarded as a hypergraph-generalization of CA on networks~\cite{propagation}. The infections through free, non-nested hyperedges, on the other hand, is approximated by means of the global mean-field as in MFA.  
Applying FA to the random nested-hypergraph model in Sec.~II, 
we take the hyperedges created in step (i) as facets; those not rewired in step (iii) as nested hyperedges; and those rewired in step (iii) as free hyperedges.

To proceed, hyperedges are classified by their sizes $s$ and their numbers of infected members $n$ for each class.  We denote 
the fraction of facets having $n$ infected nodes at time $t$ among all facets of size $s_m$ as $C_{s_m,n}(t)$; the fraction of nested (free) hyperedges where $n$ nodes are infected at time $t$ among all nested (free) hyperedges of size $s$ as $C^{(\mathrm{nested})}_{s,n}(t)$ [$C^{(\mathrm{free})}_{s,n}(t)$]; the fraction of susceptible nodes having the degree vector $\vec{\mathbf{k}}$ at time $t$ as $S_{\vec{\mathbf{k}}}(t)$. The primary quantity of interest is the  
fraction of infected nodes (i.e., infection density) at time $t$, $I(t)$, given by 
\begin{equation}
I(t)\equiv 1-\sum_{\vec{\mathbf{k}}} S_{\vec{\mathbf{k}}}(t).
\label{eq:eqqq}
\end{equation} 

In the mean-field-theoretical spirit, the main assumption behind FA is that the infected nodes are uniformly distributed within a facet, that is, $C_{s_{m},q}(t)$ acts as a local mean-field to $C^{(\mathrm{nested})}_{s,n}(t)$.
The key step in the FA is thus to approximate 
$C^{(\mathrm{nested})}_{s,n}(t)$ as follows:
\begin{equation}
\begin{array}{ll}
C^{(\mathrm{nested})}_{s,n}(t)\approx\dfrac{\sum\limits_{q=n}^{s_m} \varepsilon_s H_{s_m}{s_m \choose s}{q \choose n}{s_m-q \choose s-n}C_{s_m,q}(t)}{\varepsilon_s H_{s_m}{s_m \choose s}},
\label{eq:eqin}
\end{array}
\end{equation}
where in a facet with $q$ infected nodes there are on average ${q \choose n}{s_m-q \choose s-n}\varepsilon_s$ size-$s$ nested hyperedges with $n$ infected nodes.   
On the other hand, $C^{(\mathrm{free})}_{s,n}(t)$ is approximated by using $S_{\vec{\mathbf{k}}}(t)$ as in MFA: 
\begin{equation}
\begin{array}{ll}
C^{(\mathrm{free})}_{s,n}(t)\approx \binom{s}{n}\left(\Phi_s(t)\right)^n \left(1-\Phi_s(t)\right)^{s-n},
\label{eq:eqout}
\end{array}
\end{equation}
with $\Phi_s(t)\equiv \sum_{{\vec{\mathbf{k}}}} k^{(s,\mathrm{free})}(P_{\vec{\mathbf{k}}}-S_{\vec{\mathbf{k}}}(t))/\sum_{\vec{\mathbf{k}}} k^{(s,\mathrm{free})}P_{\vec{\mathbf{k}}}$,
the probability that a node in a random free hyperedge of size $s$ is infected at time $t$.

In the higher-order SIS dynamics, 
$S_{\vec{\mathbf{k}}}(t)$ changes over time as follows:
\begin{align}
\dfrac{dS_{\vec{\mathbf{k}}}}{dt} &=\mu\left[P_{\vec{\mathbf{k}}}-S_{\vec{\mathbf{k}}}\right]\nonumber
\\
&\quad
-\sum\limits_{n=0}^{s_{m}-1} \beta(s_{m},n)k^{(s_m)}W_{s_m,n}(t)S_{\vec{\mathbf{k}}}
\nonumber\\
&\quad
-\sum\limits_{s=2}^{s_{m}-1} \sum\limits_{n=0}^{s-1} \beta(s,n) k^{(s,\mathrm{nested})}W^{(\mathrm{nested})}_{s,n}(t)S_{\vec{\mathbf{k}}}\nonumber
\\
&\quad-\sum\limits_{s=2}^{s_{m}-1} \sum\limits_{n=0}^{s-1} \beta(s,n) k^{(s,\mathrm{free})}W^{(\mathrm{free})}_{s,n}(t)
S_{\vec{\mathbf{k}}},
\label{eq:eq1}
\end{align}
where we use the three kinetic factors: 
$(i)$
$W_{s_{m},n}(t)$, the probability that a facet of size $s_m$ including a randomly selected susceptible node has $n$ infected nodes at time $t$, which is proportional to $(s_m-n)C_{s_m,n}(t)$ and satisfies the constraint $\sum_{n=0}^{s_m} W_{s_m,n}(t)=1$, given by
\begin{align}
W_{s_{m},n}(t)=\frac{(s_m-n)C_{s_m,n}(t)}{\sum_{q=0}^{s_m} (s_m-q)C_{s_m,q}(t)}~;
\label{eq:kinefac}
\end{align}
$(ii)$
$W^{(\mathrm{nested})}_{s,n}(t)$, the probability that a nested hyperedge of size $s$ including a randomly selected  susceptible node has $n$ infected nodes at time $t$, which is proportional to $(s-n)C^{(\mathrm{nested})}_{s,n}(t)$ and satisfies the constraint $\sum_{n=0}^{s} W^{(\mathrm{nested})}_{s,n}(t)=1$, given by
\begin{align}
W^{(\mathrm{nested})}_{s,n}(t)=\sum_{q=n}^{s_m} {q \choose n}{s_m-1-q \choose s-1-n}\frac{W_{s_m,q}(t)}{{s_m-1 \choose s-1}}~;
\end{align}
$(iii)$
$W^{(\mathrm{free})}_{s,n}(t)$, the probability that a free hyperedge of size $s$ including a randomly selected susceptible node has $n$ infected nodes at time $t$, which is proportional to $(s-n)C^{(\mathrm{free})}_{s,n}(t)$ and satisfies the constraint $\sum_{n=0}^{s} W^{(\mathrm{free})}_{s,n}(t)=1$, given by
\begin{align}
W^{(\mathrm{free})}_{s,n}(t)\approx\binom{s-1}{n}(\Phi_s(t))^n (1-\Phi_s(t))^{s-1-n}~.
\end{align}
$C_{s_m,n}(t)$, on which $W_{s_m,n}(t)$ and $W^{(\mathrm{nested})}_{s,n}(t)$ depend, evolves with time as follows:
\begin{align}
\dfrac{dC_{s_{m},n}}{dt} &= \mu(n+1)C_{s_{m},n+1}-{\mu}nC_{s_{m},n}\nonumber
\\
&\quad-f_{s_{m},n}(t)C_{s_{m},n}+f_{s_{m},n-1}(t)C_{s_{m},n-1},
\label{eq:eq2}
\end{align}
where $f_{s_{m},n}(t)$ is the rate that a given facet, say $h$, having $n$ infected nodes at time $t$ has one more infected one. 
A susceptible node $v$ in $h$ can get infected from the facet itself with probability $\beta(s_m,n)$; from the nested hyperedges of size $s$ in $h$ with probability $Q(s)$ which is given by
\begin{align}
Q(s)\equiv\sum_{q=0}^{s-1}\beta(s,q)\varepsilon_s {n \choose q}{s_m-1-n \choose s-1-q}~,
\end{align}
where $\varepsilon_s{n \choose q}{s_m-1-n \choose s-1-q}$ is the expected number of nested hyperedges of size $s$ to which $q$ infected nodes and $v$ belong; 
 or with probability $\xi(t)$ from the other facets $\{h'\}$, their nested hyperedges, and free hyperedges in which $v$ participates. 
Thus $f_{s_m,n}(t)$ is written as follows:  
\begin{align}
\begin{array}{ll}
    f_{s_{m},n}(t)=(s_{m}-n)\left[\beta(s_{m},n)
    +\sum\limits_{s=2}^{s_{m}-1} Q(s)+\xi(t)\right].
    \label{eq:eq4}
\end{array}
\end{align} 
To obtain the expression for $\xi(t)$, we observe that
a susceptible node $v$ in the facet $h$, having degree vector $\vec{\mathbf{k}}$, can get infected either from $k^{(s_m)}$ other facets $\{h'\}$ with probability $\sum_{n=0} \beta(s_m,n)(k^{(s_m)}-1)W_{s_m,n}(t)$; from nested hyperedges of any size $s$ in $\{h'\}$ with probability $\sum_{q=0}^{s-1} \beta(s,q)(k^{(s_{m})}-1)(k^{(s,\mathrm{nested})}/k^{(s_m)})W^{(\mathrm{nested})}_{s,q}(t)$, where $k^{(s,\mathrm{nested})}/k^{(s_m)}$ is the expected number of hyperedges of size $s$ nested in the facet $h'$; or from free hyperedges of any size $s$ with probability $\sum_{q=0}^{s-1} \beta(s,q) k^{(s,\mathrm{free})}W^{(\mathrm{free})}_{s,q}(t)$. The probability that $v$ in $h$ has degree vector $\vec{\mathbf{k}}$ is proportional to $k^{(s_m)}S_{\vec{\mathbf{k}}}(t)$. 
Therefore, $\xi(t)$ explicitly depends on $S_{\vec{\mathbf{k}}}(t)$ and the infection probabilities from three different routes outside $h$ as follows:
\begin{widetext}
\begin{align}
\xi(t) = \dfrac{\sum\limits_{\vec{\mathbf{k}}} k^{(s_m)}S_{\vec{\mathbf{k}}}(t)\left\{\sum\limits_{n=0}^{s_{m}-1} \beta(s_{m},n)(k^{(s_{m})}-1)W_{s_{m},n}(t)+\sum\limits_{s=2}^{s_{m}-1} \sum\limits_{q=0}^{s-1} \beta(s,q)\left[(k^{(s_{m})}-1)\frac{k^{(s,\mathrm{nested})}}{k^{(s_m)}}W^{(\mathrm{nested})}_{s,q}(t)+ k^{(s,\mathrm{free})}W^{(\mathrm{free})}_{s,q}(t)\right]\right\}}{\sum\limits_{\vec{\mathbf{k}}} k^{(s_m)}S_{\vec{\mathbf{k}}}(t)}.
\label{eq:eqxi}
\end{align}
The time-evolution of $S_{\vec{\mathbf{k}}}$ and $C_{s_m,n}(t)$ in Eqs.~(\ref{eq:eq1})--(\ref{eq:eq2}) can be solved using Eqs.~(\ref{eq:eqin}), (\ref{eq:eqout}), (\ref{eq:eq4}), and (\ref{eq:eqxi}).
Let us consider the stationary state for which we use the starred variables. From Eq.~(\ref{eq:eq2}), the value of $C^*_{s_m,n}$ is given by
\begin{align}
C^{*}_{s_{m},n}=\frac{C^{*}_{s_{m},0}}{n!\mu^{n}}\prod_{q=0}^{n-1} f^{*}_{s_{m},q},
\label{eq:eq12}
\end{align}
with $n\in \{1,2,...,s_m\}$. It is a function of $\xi^{*}$ by 
Eq.~(\ref{eq:eq4}) since $C^*_{s_m,n}$ depends on $f^*_{s_m,q}$;
thus we finally yield the following self-consistent equations for $\xi^*$ and $\{S^{*}_{\vec{\mathbf{k}}}\}$:
\begin{align}
\xi^* &= \dfrac{\sum\limits_{\vec{\mathbf{k}}} k^{(s_m)}S^*_{\vec{\mathbf{k}}}\left\{\sum\limits_{n=0}^{s_{m}-1} \beta(s_{m},n)(k^{(s_{m})}-1)W^*_{s_{m},n}+\sum\limits_{s=2}^{s_{m}-1} \sum\limits_{q=0}^{s-1} \beta(s,q)\left[(k^{(s_{m})}-1)\frac{k^{(s,\mathrm{nested})}}{k^{(s_m)}}W^{*(\mathrm{nested})}_{s,q}+ k^{(s,\mathrm{free})}W^{*(\mathrm{free})}_{s,q}\right]\right\}}{\sum\limits_{\vec{\mathbf{k}}} k^{(s_m)}S^*_{\vec{\mathbf{k}}}}\equiv F_1(\xi^*,\{S^{*}_{\vec{\mathbf{k}}}\}), \nonumber
\\
&S^{*}_{\vec{\mathbf{k}}}=\dfrac{\mu P_{\vec{\mathbf{k}}}}{\mu+\sum\limits_{n=0}^{s_m-1} \beta(s_m,n) k^{(s_m)}W^{*}_{s_m,n}+\sum\limits_{s=2}^{s_m-1} \sum\limits_{q=0}^{s-1} \beta(s,q)\left[ k^{(s,\mathrm{nested})} W^{*(\mathrm{nested})}_{s,q}+k^{(s,\mathrm{free})}W^{*(\mathrm{free})}_{s,q}\right]}\equiv F_2(\xi^*,\{S^{*}_{\vec{\mathbf{k}}}\}).
\label{eq:eqself}
\end{align}
\end{widetext}
Here we introduce the simplified notations $F_{1,2}$ for the right-hand side of Eq.~(\ref{eq:eqself}). 
By solving Eq.~(\ref{eq:eqself}) for $\xi^*$ and $\{S^{*}_{\vec{\mathbf{k}}}\}$, the stationary-state infection density $I^*$ is obtained from Eq.~(\ref{eq:eqqq}).

\section{Results for random nested-hypergraphs with $s_m=3$}
In this section, we present the explicit results for the higher-order SIS model with the specific higher-order infection rates given by $\beta(s,n)=\beta_{s}\delta_{n,s-1}$ \cite{SCM} on the random nested-hypergraph model with $s_m = 3$ for varying level of nestedness.
In other words, there are only triangular and pairwise infection with rates $\beta_3$ and $\beta_2$, respectively. 
As will be shown, such a simple nested-hypergraph setting provides an intuitive clue to the nontrivial role of the hyperedge-nestedness in higher-order contagion processes. 
The stationary-state infection density $I^*$ obtained from FA 
is in good agreement with that from Monte Carlo (MC) simulations, significantly more accurate than existing analytical methods. 

\subsection{Analytical solutions for $I^*$ obtained from the FA}
For our choice of the infection rate function and hypergraph parameters, $C^{*}_{s_{m},n}$ of Eq.~(\ref{eq:eq12}) becomes as follows: 
\begin{equation}
    C^{*}_{3,n}=\frac{C^{*}_{3,0}}{n!\mu^{n}}\prod_{q=0}^{n-1} \left[(3-q)(\beta_{2}\varepsilon_{2}q+\xi^{*})+\beta_{3}\delta_{q,2}\right],
\label{eq:eq17}
\end{equation}
and the self-consistent equations for $\xi^*$ and $S^{*}_{\vec{\mathbf{k}}}$ of Eq.~(\ref{eq:eqself}) are written as follows: 
\begin{widetext}
\begin{align}
\xi^{*} &= \dfrac{\sum\limits_{\vec{\mathbf{k}}} k^{(3)}S^{*}_{\vec{\mathbf{k}}}\left[\beta_3(k^{(3)}-1) W^{*}_{3,2}+\beta_2 (k^{(3)}-1)\dfrac{k^{(2,\mathrm{nested})}}{k^{(3)}} W^{*(\mathrm{nested})}_{2,1}+\beta_2 k^{(2,\mathrm{free})} W^{*(\mathrm{free})}_{2,1}\right]}{\sum\limits_{\vec{\mathbf{k}}} k^{(3)}S^{*}_{\vec{\mathbf{k}}}}\equiv F_{1}(\xi^*,\{S^{*}_{\vec{\mathbf{k}}}\})~,
\nonumber\\
&S^{*}_{\vec{\mathbf{k}}}=\dfrac{{\mu}P_{\vec{\mathbf{k}}}}{\mu+\beta_3 k^{(3)}W^{*}_{3,2}+\beta_2 k^{(2,\mathrm{nested})} W^{*(\mathrm{nested})}_{2,1}+\beta_2 k^{(2,\mathrm{free})}W^{*(\mathrm{free})}_{2,1}}\equiv F_{2}(\xi^*,\{S^{*}_{\vec{\mathbf{k}}}\})~,
\label{eq:eq15}
\end{align}
\end{widetext}
where $W^{*}_{3,2}$, $W^{*(\mathrm{nested})}_{2,1}$, and $W^{*(\mathrm{free})}_{2,1}$ are given in turn as
\begin{align}
W^{*}_{3,2}&=\frac{C^{*}_{3,2}}{\sum\limits_{q=0}^{2} (3-q)C^{*}_{3,q}}~, \nonumber\\
W^{*(\mathrm{nested})}_{2,1}&=\frac{\sum\limits_{q=1}^{2} q(3-q)C^{*}_{3,q}}{2\sum\limits_{q=0}^{2} (3-q)C^{*}_{3,q}}~, \nonumber \\
W^{*(\mathrm{free})}_{2,1}&=1-\frac{\sum\limits_{\vec{\mathbf{k}}} (m_3+k^{(3)})S^*_{\vec{\mathbf{k}}}}{2m_3}~.
\label{eq:eq203}
\end{align}
Here $m_3$ is the average number of hyperedges of size three to which a node belongs and is defined by $m_3\equiv 3H_3/N$. 

In our random nested-hypergraph model, if a node belongs to $k^{(3)}$ facets, its $k^{(2,\mathrm{nested})}$ and $k^{(2,\mathrm{free})}$ values are sharply concentrated at the respective expectation value from Eq.~(\ref{eq:averdeg}), that is, $k^{(2,\mathrm{nested})}\approx 2k^{(3)}\varepsilon_2$ for nested hyperedges and $k^{(2,\mathrm{free})}=k^{(2,\mathrm{rewired})}\approx (1-\varepsilon_2)(k^{(3)}+m_{3})$ for free hyperedges. 
Thus we approximate $P_{\vec{\mathbf{k}}}$ as follows: 
\begin{equation}
\begin{array}{ll}
P_{\vec{\mathbf{k}}}\approx \frac{m^{k^{(3)}}_3 e^{-m_3}}{k^{(3)} !}\delta_{k^{(2,\mathrm{nested})},2k^{(3)}\varepsilon_2}\delta_{k^{(2,\mathrm{free})},(1-\varepsilon_2)(k^{(3)}+m_{3})}~,
\end{array}
\end{equation}
by using the expected values inherited from $k^{(3)}$.

\begin{figure}
\centering
\includegraphics[width=.95\linewidth]{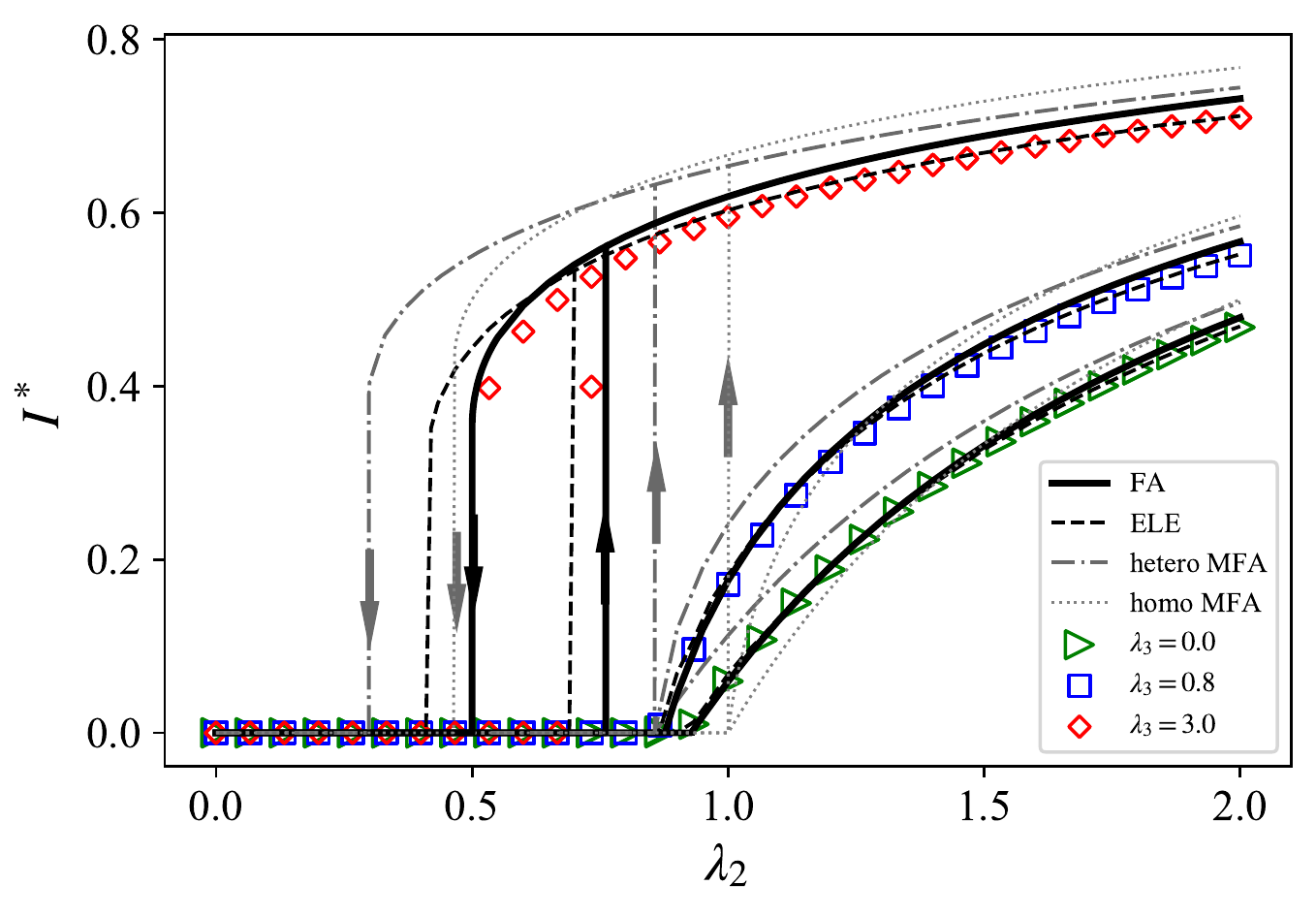}
\caption{Higher-order SIS dynamics with $\beta(s,n)=\beta_s\delta_{n,s-1}$ in fully-nested hypergraphs $(\varepsilon_2=1)$. Stationary-state infection density $I^*$ is plotted as a function of $\lambda_2\equiv m_2\beta_2/\mu$ for three different values of $\lambda_3\equiv m_3\beta_3/\mu$. Symbols depict the MC simulation results with the initial conditions of each node's infected probability $I_0=10^{-4}$ and $I_0=1$, respectively, on the random fully-nested hypergraphs of $N=10^4$ nodes with $m_3=6$. Solid lines depict the FA results, drawn for comparison together with the homogeneous MFA (labeled {`homo MFA'}), the heterogeneous MFA ({hetero MFA}), and the epidemic link equation (ELE) results.}
    \label{fig:compari}
\end{figure}

In Fig.~\ref{fig:compari}, we display the MC (symbols) and the FA (solid lines) results in the case of fully-nested ($\varepsilon_2=1$) hypergraphs for demonstration. 
Fig.~\ref{fig:compari} shows that transitions between the two phases $I^*=0$ and $I^*>0$ occur as the rescaled pairwise infectivity parameter $\lambda_2\equiv m_2\beta_2/\mu$ varies for given values of the rescaled triangular infectivity parameter $\lambda_3\equiv m_3\beta_3/\mu$.  

To demonstrate the accuracy of FA, we compare the FA results with those by existing mean-field methods. 
By neglecting the nested structure and dynamical correlations, we can make the heterogeneous and homogeneous MFA .
In the heterogeneous MFA, it is assumed that $W^*_{3,2}=\Theta^2$ and $W^{*(\mathrm{free})}_{2,1}=\Theta$; $\Theta$ is the probability   
that a node of a random hyperedge is infected in the stationary state; thus the heterogeneous MFA equation for 
$S^{*}_{\vec{\mathbf{k}}}$ of Eq.~(\ref{eq:eq15}) becomes
\begin{align}
S^{*}_{\vec{\mathbf{k}}}=\frac{{\mu}P_{\vec{\mathbf{k}}}}{\mu+\beta_{2}k^{(2)}\Theta+\beta_{3}k^{(3)}\Theta^{2}},
\label{eq:ddddd}
\end{align}
with $k^{(2)}\equiv 2k^{(3)}$.
Therefore, the self-consistency equation for $\Theta$ is obtained as follows:
\begin{align}
\Theta&=\frac{1}{m_3}\sum\limits_{\vec{\mathbf{k}}} k^{(3)}(P_{\vec{\mathbf{k}}}-S^{*}_{\vec{\mathbf{k}}})\nonumber
\\
&=\frac{1}{m_3}\sum\limits_{\vec{\mathbf{k}}} k^{(3)}P_{\vec{\mathbf{k}}}\left[\frac{\beta_{2}k^{(2)}\Theta+\beta_{3}k^{(3)}\Theta^{2}}{\mu+\beta_{2}k^{(2)}\Theta+\beta_{3}k^{(3)}\Theta^{2}}\right].
\label{eq:heteMF}
\end{align}
Note that the same set of equations can be obtained by the degree-correlated theory of Ref.~\cite{r28}.
The MFA results calculated by Eqs.~(\ref{eq:ddddd})--(\ref{eq:heteMF}) are also shown for comparison in Fig.~\ref{fig:compari}.
If we further assume $\Theta=I^{*}$ in Eq.~(\ref{eq:ddddd}), we have the homogeneous MFA equation, same as reported in Ref.~\cite{SCM}:
\begin{equation}
    \mu I^*-2\beta_{2}m_3(1-I^{*})I^{*}-\beta_{3}m_{3}(1-I^{*})(I^{*})^{2}=0.
\label{eq:eq224}
\end{equation}
The dotted lines in Fig.~\ref{fig:compari} show the solutions to Eq.~(\ref{eq:eq224}).   
As can be seen clearly in Fig.~\ref{fig:compari}, the FA is most accurate. For example, the threshold values of $\lambda_2$ at which the phase transitions occur when the initial infection density $I_0\approx 0$ do not depend on $\lambda_3$ in both MFA results but do in the MC and the FA results. 
Note that the bistable regime, where the two solutions $I^*=0$ and $I^*>0$ are stable and accessible, appears when the value of $\lambda_3$ is large enough.
The arrows pointing upward (downward) represent the phase transitions when $I_0\approx 0$ ($I_0=1$) in all the three methods. 
For fully-nested structure $(\varepsilon_2=1)$, the method of epidemic link equation with closure~\cite{r29} can be used. Its results are plotted together in Fig.~2 for comparison. 

\begin{figure}[t]
    \centering
    \includegraphics[width=.95\linewidth]{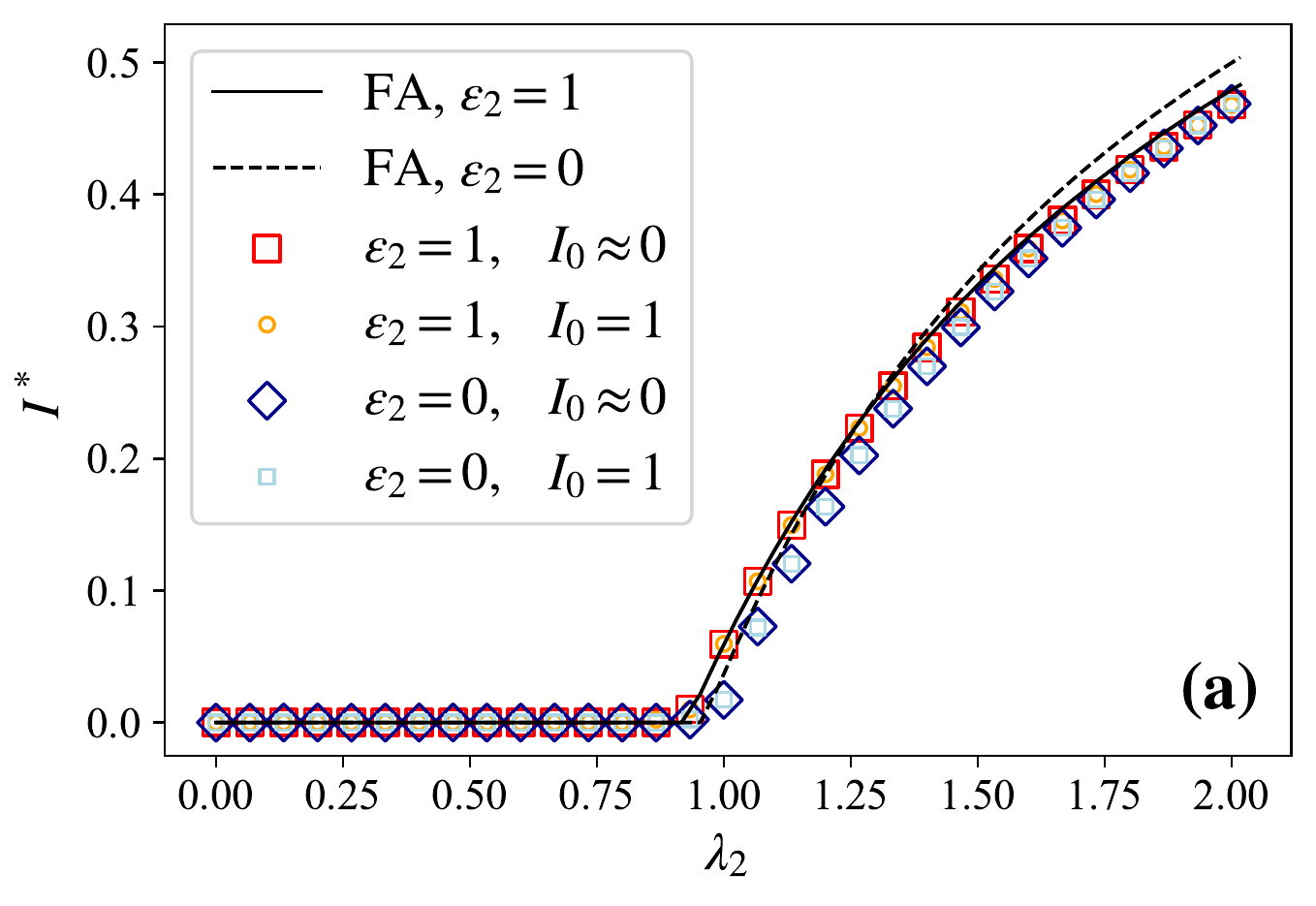}
    \includegraphics[width=.95\linewidth]{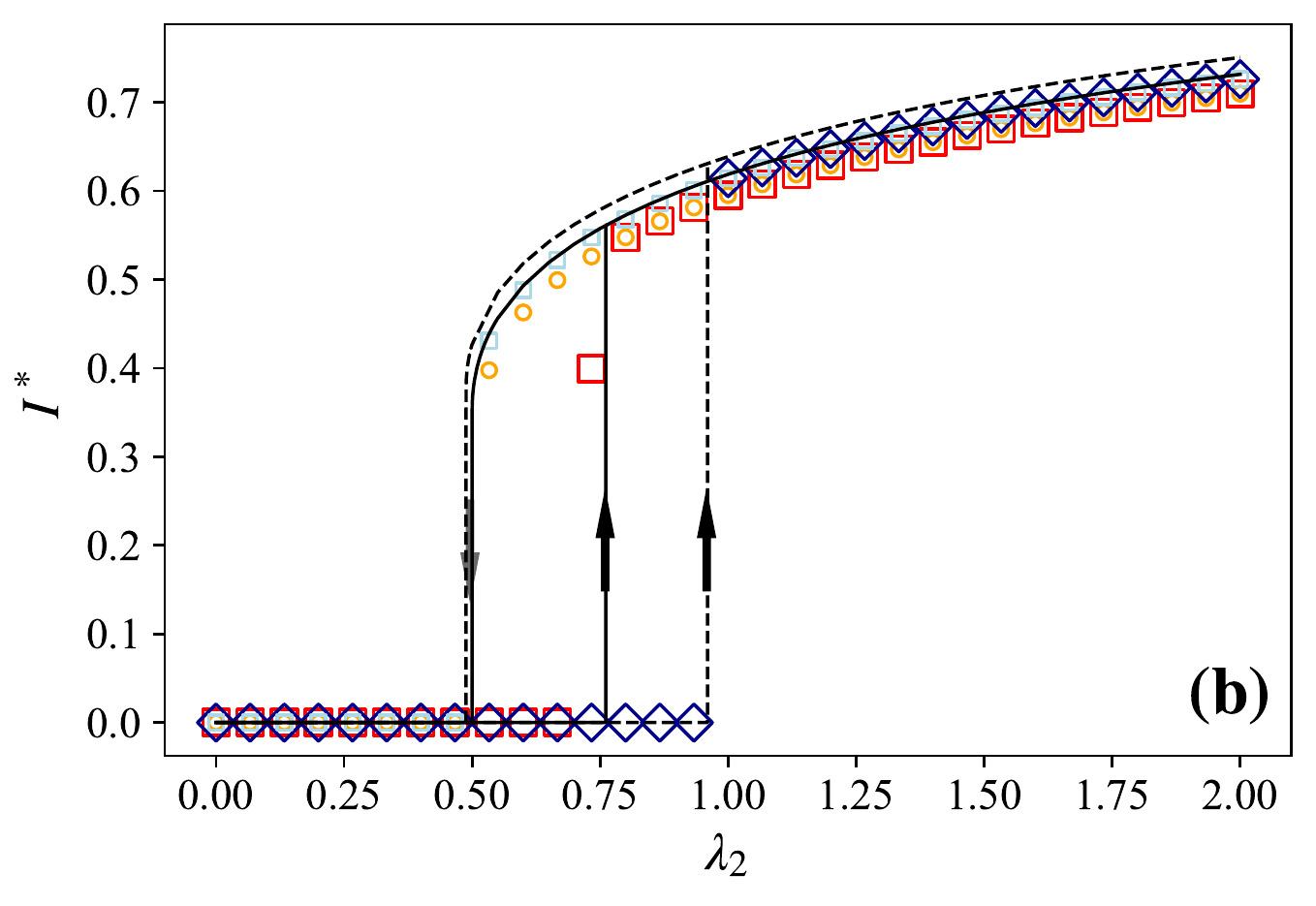}
    \caption{Comparison of the stationary-state infection density $I^*$ on fully-nested ($\varepsilon_2=1$) and non-nested ($\varepsilon_2=0$) hypergraphs consisting of $N=10^4$ nodes with $m_3=6$ for (a) $\lambda_3=0$ and (b) $\lambda_3=3$. The larger (smaller) symbols portray MC simulations with $I_0=0.0001$ ($I_0=1$).}
\label{fig:compar}
\end{figure}

Using the FA, we compare the contagion dynamics on fully-nested ($\varepsilon_2=1$) and non-nested ($\varepsilon_2=0$) hypergraphs in Fig.~\ref{fig:compar}. There exists bistable regime for large enough $\lambda_3$ for both $\varepsilon_2=0$ and $\varepsilon_2=1$. Yet, the bistable regime is broader in non-nested hypergraphs than fully-nested ones. We can find that the fixed point $I^*=0$ becomes unstable when $\lambda_2>\lambda_{ic}$, and $I^*>0$ becomes unstable when $\lambda_2<\lambda_{pc}$; this defines the two transition points $\lambda_{ic}$ and $\lambda_{pc}$, called the invasion threshold and the persistent threshold, respectively \cite{universal}. 
In Fig.~\ref{fig:compar}(a), for $\lambda_3=0$, the transitions between $I^*=0$ and $I^*>0$ are continuous, i.e., $\lambda_{ic}=\lambda_{pc}$. On the other hand, for sufficiently large $\lambda_3$, 
discontinuous transitions are shown in Fig.~\ref{fig:compar}(b) where $\lambda_{ic}>\lambda_{pc}$. 
It is noteworthy that the $\lambda_3$-dependent invasion threshold $\lambda_{ic}$ in higher-order contagion processes has been noted also in Refs.~\cite{cca,st2022influential}.

\subsection{Phase diagram}
\begin{figure}    
\centering
    \includegraphics[width=.95\linewidth]{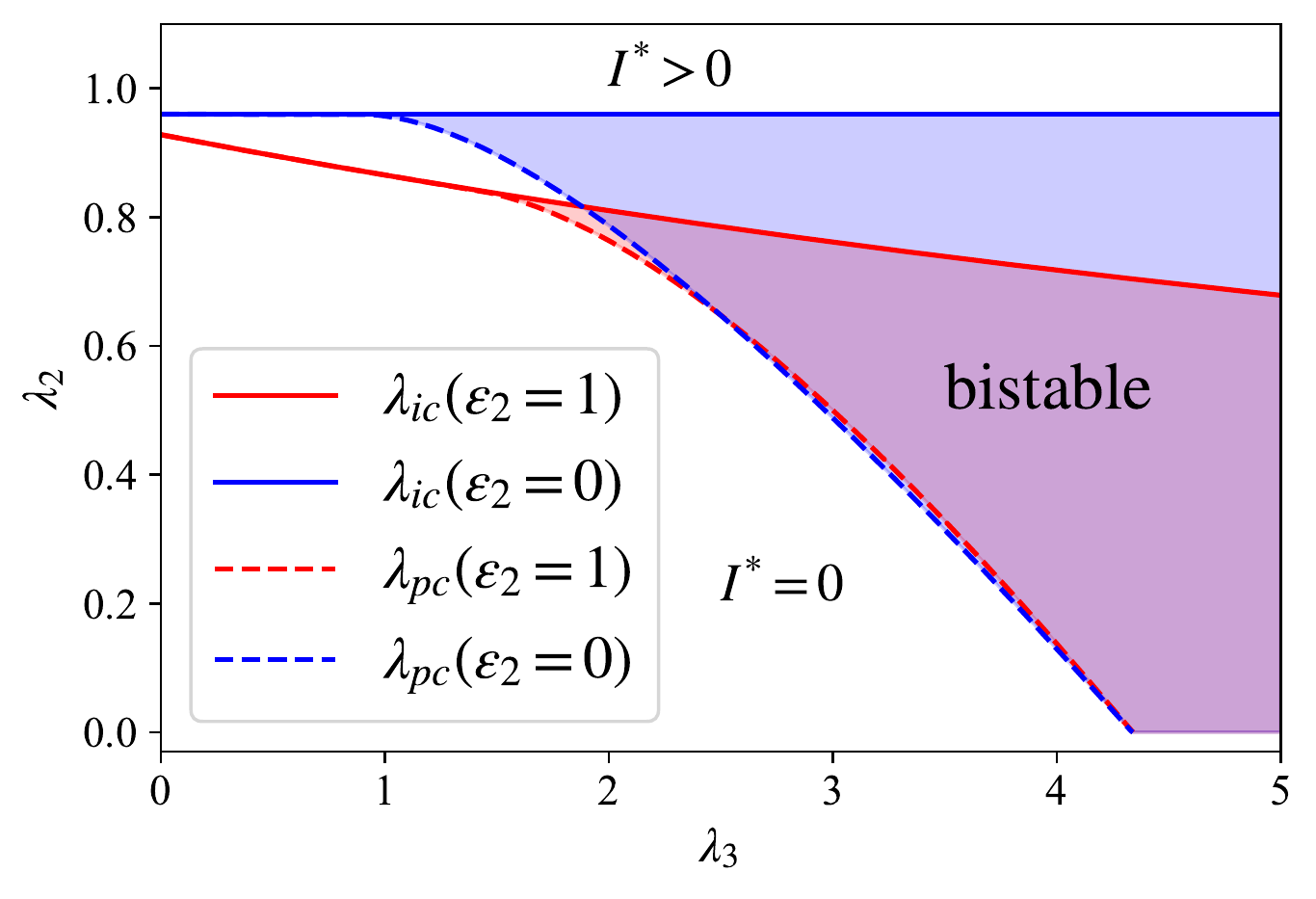}
    \caption{Phase diagram from the FA for fully-nested (red) and non-nested (blue) hypergraphs with $m_3=6$. The solid (dashed) lines represent the values of the invasion threshold $\lambda_{ic}$ (the persistent threshold $\lambda_{pc}$). In the endemic (disease-free) regime where $\lambda_2>\lambda_{ic}$ ($\lambda_2<\lambda_{pc}$), the only stable solution is $I^*>0$ ($I^*=0$). The shaded area where $\lambda_{pc}<\lambda_2<\lambda_{ic}$ corresponds to the bistable regime.}
    \label{fig:Fig3}
\end{figure}

In Fig.~\ref{fig:Fig3}, we show the phase diagram 
constructed by the FA. Phase boundaries are obtained by using Eq.~(\ref{eq:eq15}), determining the values of $\lambda_{ic}$ and $\lambda_{pc}$ for given $\lambda_3$.   
The value of $\lambda_{ic}$ is determined by the following condition:
\begin{equation}
    \left.{\frac{\operatorname{d}\!F_1}{\operatorname{d}\!\xi^{*}}}\right|_{\xi^{*}=0}=1.
    \label{eq:eq24}
\end{equation}
Note that the invasion threshold can be derived analytically in the two extreme cases, $\varepsilon_2=0$ and $\varepsilon_2=1$.  
The first derivative value of $C^{*}_{3,n}$ is given by
\begin{align}
    \left.{\frac{\operatorname{d}\!C^{*}_{3,n}}{\operatorname{d}\!\xi^{*}}}\right|_{\xi^{*}=0}=
    (n-1)!{3 \choose n}\frac{\varepsilon_2^{n-1}\lambda_{ic}^{n-1}}{\mu m_2^{n-1}}+\frac{2\lambda_{ic}\lambda_3\varepsilon_2\delta_{n,3}}{\mu m_2^2}.
\label{eq:eq26}
\end{align} 
In Eqs.~(\ref{eq:eq17})--(\ref{eq:eq203}), we can see that $C^*_{3,n}=\delta_{n,0}$ and $S^{*}_{\vec{\mathbf{k}}}=P_{\vec{\mathbf{k}}}$ for $\xi^*=0$. 
Thus, Eq.~(\ref{eq:eq24}) becomes:
\begin{align}
\left.{\dfrac{\operatorname{d}\!F_1}{\operatorname{d}\!\xi^{*}}}\right|_{\xi^*=0}&=
\dfrac{\lambda_3\lambda_{ic}\varepsilon_2}{m_2}
+\lambda_{ic}\varepsilon_2+\dfrac{\lambda_{ic}^2 \varepsilon_2^2}{m_2}\nonumber
\\
&\quad+\dfrac{\mu\lambda_{ic}(m_2+1)(1-\varepsilon_2)}{m_2}A(m_2,\varepsilon_2,\lambda_{ic},\lambda_{3})\nonumber
\\
&=1,
\label{eq:eq27}
\end{align}
where $A(m_2,\varepsilon_2,\lambda_{ic},\lambda_3)\equiv \left.{\frac{\operatorname{d}\!W^{*(\mathrm{free})}_{2,1}}{\operatorname{d}\!\xi^{*}}}\right|_{\xi^*=0}$, satisfying
\begin{align}
A(& m_2,\varepsilon_2,\lambda_{ic},\lambda_{3})=\nonumber \\
&\frac{\lambda_{ic}\lambda_3 \varepsilon_2(m_2+1)}{\mu m^2_2}
+\frac{(m_2+1)}{\mu}\left[\frac{\lambda_{ic}\varepsilon_2}{m_2} 
+\frac{\lambda^2_{ic}\varepsilon^2_2}{m^2_2}\right] \nonumber\\
&+\frac{(1-\varepsilon_2)(2m_2+1)\lambda_{ic}}{2m_2}A(m_2,\varepsilon_2,\lambda_{ic},\lambda_3)~.
\label{eq:eqA}
\end{align}

Transition points constituting the phase boundaries  thus obtained are: For non-nested ($\varepsilon_2=0$) hypergraphs, we obtain
\begin{align}
\lambda_{ic}&=\frac{2m_2}{(2m_2+1)}\nonumber
\\
&=\frac{4m_3}{4m_3+1}
\label{eq:eq28}
\end{align}
On the other hand, for fully-nested hypergraphs ($\varepsilon_2=1$), we obtain from Eq. (\ref{eq:eq27})
\begin{equation}
\begin{array}{ll}
    \lambda_{ic}=\dfrac{-2m_3-\lambda_3+\sqrt{4m_{3}^{2}+\lambda_{3}^{2}+4\lambda_{3}m_3+8m_3}}{2},
\label{eq:eq29}
\end{array}
\end{equation}
which is a monotonically decreasing function of $\lambda_{3}$. The smaller values of $\lambda_{ic}$ with $\varepsilon_2=1$ than with $\varepsilon_2=0$ seen in
the solid lines in Fig.~\ref{fig:Fig3} given from Eqs.~(\ref{eq:eq28})--(\ref{eq:eq29}) show that infectious diseases spread more easily over the fully-nested hypergraphs than the non-nested ones. To understand the origin, consider an early stage of epidemic spreading with $I_{0}\approx0$. At this stage, most infection arises via pairwise interactions. Once a hyperedge of size $2$ is fully infected, the condition for triangular infection through the hyperedge of size $3$ is met concomitantly in the case of $\varepsilon_{2}=1$, whereas it is not always so in the case of $\varepsilon_2=0$.  

On the contrary, if the value of $I_0$ is large enough, the condition for triangular-infection processes is easily met in both cases of $\varepsilon_{2}=1$ and $\varepsilon_{2}=0$. The dashed lines in Fig.~\ref{fig:Fig3}, obtained numerically by
\begin{equation}
\left.{\frac{\operatorname{d}\!F_1}{\operatorname{d}\!\xi^{*}}}\right|_{\xi^{*}>0}=1,
\end{equation}
demonstrate that the values of $\lambda_{pc}$ for fully-nested hypergraphs and non-nested hypergraphs are different only slightly, compared to the difference between the invasion threshold lines. 

\begin{figure}    
\centering
    \includegraphics[width=.95\linewidth]{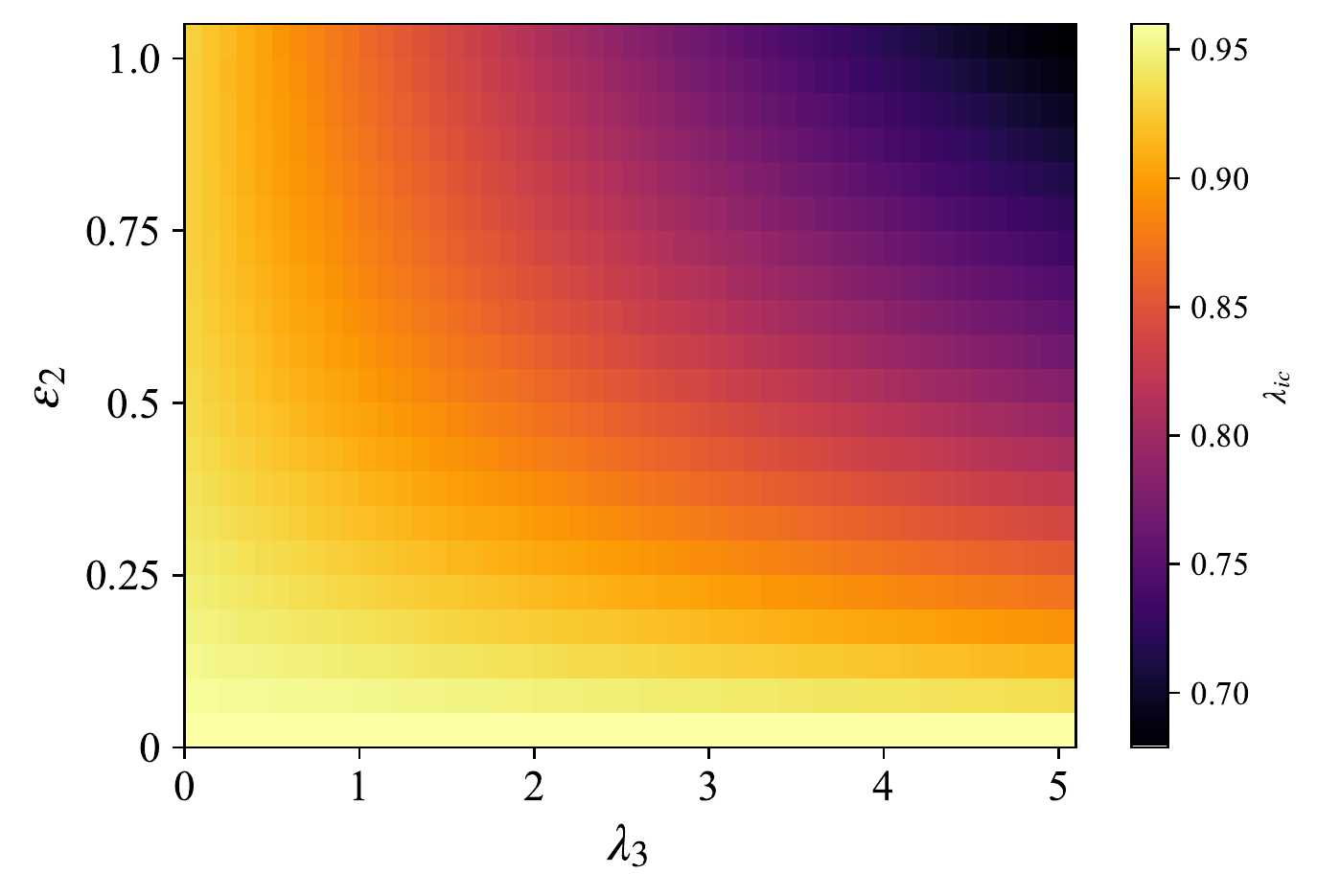}
    \caption{Invasion threshold $\lambda_{ic}$ for the entire range of nestedness parameter $0\le\varepsilon_2\le1$ obtained by FA. Hypergraph model parameters used are the same as in Fig.~4.}
    \label{fig:Fig_invasion}
\end{figure}

Finally, we computed the invasion threshold $\lambda_{ic}$ in the entire range of nestedness parameter $\varepsilon_2$, shown in Fig.~\ref{fig:Fig_invasion}. We found that the invasion threshold $\lambda_{ic}$ decreases with $\lambda_3$ for generic nestedness parameter $\varepsilon_2>0$, and the effect gets more pronounced with increasing $\varepsilon_2$.

\section{Conclusion and Discussion} 
In this paper, we have studied the effect of hyperedge-nestedness on higher-order contagion dynamics.
To this end, we have introduced and formulated the FA framework for the higher-order SIS process on the proposed model of random nested-hypergraphs with tunable nestedness. 
The mathematical formulation allows us to capture how the hyperedge-nestedness facilitates the higher-order contagion more accurately than existing methods.

By applying the FA formalism to the higher-order SIS model with $\beta(s,n)=\beta_{s}\delta_{n,s-1}$ on the nested hypergraphs composed of pairwise and triangular interactions, we compared the phase boundaries given from the FA for fully-nested and non-nested hypergraphs. We  
found that infectious diseases can spread over the fully-nested hypergraphs with lower pairwise infectivity. In other words, an increase in the hyperedge-nestedness makes a drop in the invasion threshold by promoting triangular infections. 
The FA also forecasts the appearance of the bistable regime for large enough $\lambda_3$ on both fully-nested and non-nested hypergraphs. MC simulations support these predictions, indicating the usefulness of our analytic method.  

As future works, more comprehensive investigation of the nested-hypergraph structures such as the model with larger $s_m>3$ or even with heterogeneous $s_m$ may reveal extra emergent effects of nestedness in addition to its generic effects identified in this paper.
Another direction of immediate interest is to apply FA to real-world hypergraphs with nestedness~\cite{motif}. Random nested-hypergraph modeling of the real-world hypergraph is an exciting open problem~\cite{recon}. 
Furthermore, developing the FA to incorporate additional higher-order correlation structures omnipresent in real-world hypergraphs such as higher-order components ~\cite{jungho} would also be desirable towards this goal.
Finally,
we anticipate that the FA framework could be utilized as a tool for studying other higher-order contagion processes which may be more complex than the higher-order SIS model such as susceptible-infectious-recovered and susceptible-exposed-infectious-recovered dynamics on nested hypergraphs. 

\begin{acknowledgments}
We thank Y. Yang and D. Rho for discussion in the preliminary stage of project.
This work was supported in part by the National Research Foundation of Korea (NRF) grant funded by the Korea government (MSIT) (No. 2020R1A2C2003669) (K.-I.G.) and by a KIAS Individual Grant No.CG079901 at Korea Institute for Advanced Study (D.-S.L.).
\end{acknowledgments}
\nocite{*}

\bibliography{reflist}

\end{document}